\documentclass[%
 reprint,
 bibnotes,
 amsmath,amssymb,
 aps
]{revtex4-1}

\usepackage{graphicx}
\usepackage{epstopdf}
\usepackage{dcolumn}
\usepackage{bm}
\usepackage{hyperref}

\begin{document}
\title{New method for non-standard invisible particle searches in tau lepton decays}
\author{E. De La Cruz-Burelo}
\email[Email address: ]{e.delacruz.burelo@cinvestav.mx}
\author{A. De Yta-Hernandez}
\email[Email address: ]{adeyta@fis.cinvestav.mx}
\affiliation{Centro de Investigacion y de Estudios Avanzados del Instituto Politecnico Nacional, Av. IPN 2508, San Pedro Zacatenco, Mexico City, 07360}
\author{M. Hernandez-Villanueva}
\email[Email address: ]{mhernan7@olemiss.edu}
\affiliation{University of Mississippi, University, Mississippi 38677.}

\date{\today} 

\begin{abstract}
Motivated by models proposed to explain Standard Model anomalies, 
and the unprecedented $\tau^{+}\tau^{-}$ data to be collected by the Belle~II experiment during the next years, we study the kinematics of tau pair decays and propose a new method to search for lepton flavor violating processes in tau lepton decays to invisible beyond Standard Model particles, such as $\tau \to \ell  \alpha$, where $\ell$ is either an electron or a muon, and $\alpha$ is a massive particle that escapes undetected. The new method improves by one order of magnitude the expected upper limit on the $\tau \to \ell \alpha$ production in 3x1 prong tau decays and establishes the possibility of performing this search in 1x1 prong tau decays which has not been previously considered.
\end{abstract}

\maketitle

\section{Introduction}

The Standard Model of particles physics (SM) has been incredibly successful in explaining all observed data up today, with few remaining tensions between prediction and experiment, for instance, the long-standing 3.7 standard deviations discrepancy in the anomalous magnetic moment of the muon $a_{\mu} = (g-2)_{\mu}/2$ \cite{Bennett:2006fi, Hagiwara:2011af, Keshavarzi:2019abf, Davier:2019can,Aoyama:2020ynm}. However, observed phenomena as the predominance of matter over antimatter in the universe, the neutrino masses, or dark matter, among others, suggest physics beyond the SM (BSM). In consequence, searching for new physics has become of primary importance, and without clear indications of the SM applicability boundaries, a broad exploration strategy needs to be followed. One of these strategies involves searching for the extremely SM suppressed Lepton Flavour Violating (LFV) processes, which observation will be a clear signal of BSM physics. 

In the search for LFV processes, the tau lepton is a unique laboratory with an indirect probe to energies not directly accessible by accelerators. Of particular importance are LFV tau lepton decays to invisible BSM particles produced in various models containing axion-like particles~\cite{Davidson:1981zd,Wilczek:1982rv, Berezhiani:1989fp,Calibbi:2020jvd} or new $Z^{\prime}$ gauge bosons~\cite{Altmannshofer:2016jzy,Altmannshofer:2016brv,Heeck:2016xkh}, and which aim to explain SM anomalies like the $a_{\mu}$ discrepancy. One possibility of such processes is $\tau \to \ell \alpha$, where $\ell$ is either an electron or a muon, and $\alpha$ is a massive particle that escapes to detection. This decay appears in several new physics models~\cite{Altmannshofer:2016brv,Heeck:2016xkh,Grinstein:1985rt, Berezhiani:1989fp, Feng:1997tn,Asai:2018ocx} and are of interest not only due to the $a_{\mu}$ deviation, but also because very light particles could serve as dark matter candidates \cite{Fayet:2004bw, Pospelov:2007mp}, or to answer the proton radius puzzle~\cite{Pohl:2013yb}. 

Tau LFV processes will be searched in the upcoming data from the Belle II experiment~\cite{Abe:2010sj, Kou:2018nap}, where an unprecedented statistics of $\sim5\times 10^{10}$ tau lepton pairs is expected. However, it will take several years for Belle II to accumulate the data necessary to improve the current exclusion limits on the BSM processes accessible to the tau lepton sector~\cite{Villanueva:2018pbk, Konno:2020tmf}. Therefore, new methods with superior statistical performance than the standard searching techniques could expand the data output, and this is the aim of the present work.

Inspired by searches for dark matter or invisible heavy particles in $X\bar{X}\to (Y_a + N)(Y_b + \bar{N})$ processes \cite{HarlandLang:2012gn,Xiang:2016jni,Christensen:2014yya}, with $Y_a$ and $Y_b$ being the only detectable products, we generalize the idea to $X\bar{X}\to (\sum_{i=1}^{n}Y_{ai} + N_{1})(\sum_{j=1}^{m}Y_{bj} + N_2)$ decays where $Y$ represents visible particles and $N$ particles that evades detection. 
This generalization allows us to study $X\bar{X}$ pair decays with BSM processes in one decay, and SM processes with a missing particle in the complementary decay; hence increasing the possibility of a non-SM particle production compared to requiring a double creation of the unknown particle as in previous studies. 

From the generalized case of the $X\bar{X}$ pair decay, we determine a kinematic constraint that relates the masses of the mother particle $X$ and the undetectable particles $N_{1}$ and $N_{2}$. We use this relationship to propose new searching variables for non-standard invisible particles in tau lepton decays from collisions with initial state energy and momentum well defined. 
B-Factories such as BaBar, Belle, and Belle II provide an ideal environment with these characteristics, colliding electrons and positrons with a known energy in the center-of-mass system (CMS). 

We apply our findings to the search for LFV processes $\tau \to \ell \alpha$ in simulated 3x1 prong decays data emulating the Belle II experiment conditions. We propose a new two-dimensional method that, compared to the standard search technique, reduces an order of magnitude the expected upper limit on the production of this BSM decay, and opens the possibility of an additional search in 1x1 prong decays, which is most abundantly produced. 

\section{Kinematic constraints}\label{KConstraints}

\begin{figure}
\includegraphics[width=8cm]{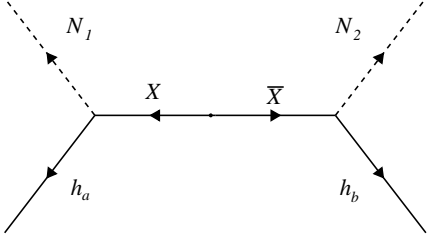}
\caption{$X\bar{X}$ production topology in the center-of-mass frame. Each $X$ decays to a detectable product $h$ and an invisible particle $N$ that escapes undetected. \label{diagram}}
\end{figure}

Let us consider the pair decay 
\begin{equation}
X\bar{X}\to \Big(\sum_{i=1}^{n}Y_{ai} + N_{1}\Big)\Big(\sum_{j=1}^{m}Y_{bj} + N_2\Big)
\end{equation}
at CMS energy $\sqrt{s}$. 
Here $N_1$ and $N_2$ are invisible particles that elude detection, and $Y_{ai}$ and $Y_{bj}$ are
the $i$-th and $j$-th visible particles from the $X$ and $\bar{X}$ decays, respectively. 
To facilitate calculations, we use $h_a$ and $h_b$ to indicate $\sum_{i=0}^{n} Y_{ai}$ and $\sum_{i=0}^{m} Y_{bj}$. Then the $X\bar{X}$ pair decay is treated as illustrated in Fig.~\ref{diagram}, 
with $p_a=(E_a,{\bf p}_a)$, $p_b=(E_b,{\bf p}_b)$, $p_1=(E_{1},{\bf p}_1)$, and $p_2=(E_{2},{\bf p}_2)$ being the four-momenta at CMS for $h_a$, $h_b$, $N_{1}$, and $N_{2}$, respectively. 
We follow a similar approach as in Ref. \cite{Xiang:2016jni}, but allowing both decays to produce different missing particles. The kinematic equations for the process are
\begin{eqnarray}
 q^{\mu} & = & p_a^{\mu} + p_b^{\mu} + p_1^{\mu} + p_2^{\mu}, \;\; \mu=0,1,2,3, \label{kinEq1}\\
 p_1^2 & = & m_1^2 \label{kinEq2},\\
 p_2^2 & = & m_2^2 \label{kinEq3},\\  
 (p_a + p_1)^2 &=& (p_b + p_2)^2 = m_{X}^2 ,\label{kinEq4}
\end{eqnarray}
where $m_1$, $m_2$, and $m_{X}$ are the masses of $N_1$, $N_2$, and $X$, respectively.  
By defining the normalized variables $\mu_{i} = m_{i}/\sqrt{s}$, $z_{i}=E_{i}/\sqrt{s}$, ${\bf a}={\bf p}_{a} /\sqrt{s}$, ${\bf b}={\bf p}_{b} /\sqrt{s}$, ${\bf k}_1={\bf p}_{1} /\sqrt{s}$, ${\bf k}_2={\bf p}_{2} /\sqrt{s}$, from Eq. \ref{kinEq1} we have ${\bf k}_{1} + {\bf k}_{2} + {\bf a} + {\bf b} = {\bf 0}$ and $z_{1} + z_{2} + z_{a} + z_{b} = 1$. Then we can rewrite Eqs. \ref{kinEq2}--\ref{kinEq4} as
\begin{eqnarray}
 |{\bf k}_{1}|^2 + \mu_{1}^{2} & = & z_{1}^{2} \label{kin1Eq1},\\
 |{\bf k}_{1} + {\bf a} + {\bf b}|^2 + \mu_{2}^{2} & = & (1 - z_{a} - z_{b} - z_{1})^{2}, \label{kin1Eq2}\\ 
 |{\bf k}_{1} + {\bf a}|^2 + \mu_{X}^{2} & = & \frac{1}{4}, \label{kin1Eq3}
\end{eqnarray}
where we use $z_{X}=1/2$. From Eq. \ref{kin1Eq1} we have
\begin{equation}\label{Eq1}
{\bf k}_1\cdot {\bf k}_1 = k^2_1 = \Big( \frac{1}{2} - z_{a} \Big)^{2} - \mu_{1}^{2},
\end{equation}
and from Eq. \ref{kin1Eq2} and Eq. \ref{kin1Eq3} we obtain
\begin{equation}\label{Eq2}
 {\bf a}\cdot {\bf k}_1 = A, 
\end{equation}
and
\begin{equation} \label{Eq3}
 {\bf b}\cdot {\bf k}_1 = B, 
\end{equation}
where
\begin{eqnarray}
A &=&\frac{1}{2}\Big( z_a - z_a^2 - \mu_{X}^{2} + \mu_{1}^{2} - |{\bf a}|^2 \Big),\label{EqA}\\
B &=&\frac{1}{2}\Big( z_b^2 - z_b + \mu_{X}^{2} - \mu_{2}^{2} - |{\bf b}|^2 \Big) - {\bf a}\cdot {\bf b},\label{EqB}
\end{eqnarray}
In addition, ${\bf k}_1$, ${\bf a}$, and ${\bf b}$, must comply with
\begin{eqnarray}
 |{\bf k}_1\times {\bf a}\times {\bf b}|^2 &=& | ({\bf b}\cdot {\bf k}_1) {\bf a} - ({\bf a}\cdot {\bf k}_1){\bf b} |^2 ,\nonumber\\
 &=& |{\bf k}_1|^2 |{\bf a}\times {\bf b}|^2 \sin^{2}{\theta}, \nonumber\\
 &\leq& |{\bf k}_1|^2 |{\bf a}\times {\bf b}|^2,
\end{eqnarray}
and by using Eqs. \ref{Eq1}--\ref{Eq3}, this transforms to
\begin{equation}\label{EqDes}
 |B{\bf a} - A {\bf b}|^2 - \Big[\Big( \frac{1}{2} - z_{a} \Big)^{2} - \mu_{1}^{2}\Big]|{\bf a}\times {\bf b}|^2\leq 0,
\end{equation}
From Eqs. \ref{EqA} and \ref{EqB}, it is straightforward to show that
\begin{equation}
    B{\bf a} - A{\bf b} = \frac{1}{2}\Big((\mu_{X}^2 - \mu_{2}^{2}){\bf a} + (\mu_{X}^2 - \mu_{1}^{2}){\bf b} + {\bf H}\Big),
\end{equation}
where 
\begin{equation}
{\bf H} \equiv \Big( z_b^2 - z_b - |{\bf b}|^2 - 2{\bf a}\cdot {\bf b}  \Big) {\bf a} + \Big( z_a^2 - z_a + |{\bf a}|^2 \Big) {\bf b}.    
\end{equation}
Then Eq. \ref{EqDes} transforms to
\begin{eqnarray}\label{EqMaster}
 A_1(\mu_{X}^2 - \mu_{1}^2)^2 + A_2(\mu_{X}^2 - \mu_{2}^2)^2  + \nonumber\\
 A_3(\mu_{X}^2 - \mu_{1}^2)(\mu_{X}^2 - \mu_{2}^2)  + \nonumber \\ 
 B_1(\mu_{X}^2 - \mu_{1}^2) + B_2(\mu_{X}^2 - \mu_{2}^2) + \nonumber\\
 C_1\mu_{1}^2 + D_1 \leq 0, 
\end{eqnarray}
where
\begin{eqnarray}
A_1 &=& |{\bf b}|^2,\\
A_2 &=& |{\bf a}|^2,\\
A_3 &=& 2({\bf a}\cdot{\bf b}),\\
B_1 &=& 2({\bf b}\cdot{\bf H}),\\
B_2 &=& 2({\bf a}\cdot{\bf H}),\\
C_1 &=& 4|{\bf a}\times{\bf b}|^2,\\
D_1 &=& {\bf H}\cdot {\bf H} -   4|{\bf a}\times{\bf b}|^2\Big( \frac{1}{2} - z_{a} \Big)^{2}.
\end{eqnarray}
Equation \ref{EqMaster} is our main result and contains the available kinematics information of the $X\bar{X} \to (h_{a} + N_{1}) (h_{b} + N_{2})$ decay.

\section{Search for $\tau\to \ell\alpha$ decays}

The last search for $\tau\to \ell\alpha$ decays was performed by the ARGUS Collaboration~\cite{Albrecht:1995ht} in $\tau \to \ell + anything$ data, with  $\ell$ being an electron or a muon. The main challenge in the $\tau\to \ell  \alpha$ search is to separate these signal decays from the same-signature SM process $\tau\to \ell  \bar{\nu}_{\ell} \nu_{\tau}$. Since the signal is a two-body decay, ARGUS used the fact that, in contrast to the three-body SM process, in the tau rest frame the lepton momentum is a constant value given by
\begin{equation}\label{xrf_fix}
 b(m_{\alpha}) = \frac{m_{\tau}^2 - m_{\alpha}^2 + m_{\ell}^2}{m_{\tau}^2}
\end{equation}
where $m_{\tau}$ is the mass of the tau; $m_{\alpha}$ the mass of the $\alpha$ particle; and $m_{\ell}$ the mass of lepton. 
This feature would allow us to separate the two processes and determine $m_{\alpha}$ if the reconstruction of the tau rest frame were possible. Unfortunately, each tau decay involves a missing particle, making impossible a full reconstruction of the tau. To solve this problem, ARGUS required the other tau in the $\tau^{+}\tau^{-}$ production to decay to three pions and developed the so-called pseudo-rest-frame technique in which the lepton in the one-prong side is boosted to the tau rest frame by approximating: 
a) the one-prong side tau momentum direction as the opposite direction of the momentum of the three pions in the three-prong side;
and b) the tau energy by $E_{\tau}=\sqrt{s}/2$. From now on, we will refer to these approximations as the ARGUS method.

By using Eq. \ref{EqMaster}, we can construct other methods to search for $\tau\to \ell\alpha$ decays. For simplicity, and in order to compare to the ARGUS method, let us consider the process $(\tau^{+}\to  \pi^{+}\pi^{-}\pi^{+}\bar{\nu}_{\tau})(\tau^{-}\to e^{-} \alpha)$. 
For the decays studied in Section \ref{KConstraints}, this is a particular case where $\mu_1=\mu_{\alpha}$, $\mu_2=\mu_{\nu_{\tau}}$ and $\mu_{X}=\mu_{\tau}$. Assuming $\mu_{\nu_{\tau}}=0$, Eq. \ref{EqMaster} reduces to 
\begin{equation}\label{Eq1x1Inv}
    A_0(\mu_{\alpha}^2)^2 + B_0 \mu_{\alpha}^2 + C_0 \leq 0 ,
\end{equation}
where
\begin{eqnarray}
A_0 &=& A_1,\\
B_0 &=& -B_1 + C_1 - (2 A_1 + A_3) \mu_{\tau}^2, \\
C_0 &=& (A_1 + A_2 + A_3) \mu_{\tau}^4 + (B_1 + B_2) \mu_{\tau}^2 + D_1.
\end{eqnarray}

Then, Eq. \ref{Eq1x1Inv} translates to
\begin{eqnarray}\label{EqboundsInv}
M_{min}^2 \leq m_{\alpha}^2\leq M_{max}^2,
\end{eqnarray}
where
\begin{eqnarray}
M_{min}^2&=&(\sqrt{s})^2\Big( \frac{-B_0 - \sqrt{B_0^2 - 4A_0 C_0}}{2A_0}\Big),\label{MminZero}\\
M_{max}^2&=&(\sqrt{s})^2\Big(\frac{-B_0 + \sqrt{B_0^2 - 4A_0 C_0}}{2A_0} \Big).\label{MmaxZero}
\end{eqnarray}

According to Eq. \ref{EqboundsInv}, the distribution of the square value of these new variables, $M_{min}$ and $M_{max}$, must show endpoints at the value of $m_{\alpha}^{2}$. We can use these endpoints to untangle $\tau\to \ell \alpha $ decays from the SM processes and measure the mass of the $ \alpha $ particle in case of observation. Also, if these new variables are not highly correlated, they could be combined in a two-dimensional distribution to increase the statistical discrimination power of the method. In the rest of the paper, we will refer to these as the $M_{min}$, $M_{max}$, and 2D methods, respectively.

\subsection{Simulated data selection} \label{Selection}
 
To study the kinematic bounds in Eq. \ref{EqboundsInv} at the energies of the Belle II experiment, we simulate $e^{+}e^{-}\to \tau^{+}\tau^{-}$ and $e^{+}e^{-}\to q\bar{q}$ processes at  $\sqrt{s}=10.58$ GeV using Pythia8 \cite{Sjostrand:2014zea} implemented in ROOT 6.20 \cite{Brun:1997pa}.
To account for $\tau\to e \alpha $ decays, we added to Pythia8 a new stable $ \alpha $ spin-0 particle, which decay is simulated using a phase-space model.
We estimate the number of simulated events for $\tau^{+}\tau^{-}$ and $q\bar{q}$ decaying to SM processes from the cross-sections reported by Belle II \cite{Kou:2018nap}. 
For particles with transverse momentum $p_{T}$, the momentum precision $\sigma$ in the Belle II experiment \cite{Bertacchi:2020eez}, varies from $\sigma/p_{T}\approx 5$\% for very low $p_T$ particles, to  $0.3$\% for $p_T\geq 0.5$ GeV.
To have more realistic simulated data, we apply Gaussian smearing to the momentum components of the final state particles for an average precision of $\sigma_{p_{T}}/p_{T} = 1$\%. 

In order to select 3x1 prong tau decays, we require per event four charged particles in the final state, and no more than one photon with an energy higher than 0.05 GeV; the latter to suppress decays with neutral pions decaying to photons in the kinematic regime of the photon reconstruction in the Belle II detector. 
Tau pair decays are produced back-to-back in the CMS and their decay produce jet-like events, with two cones of collimated particles around the thrust axis ${\bf n}_{T}$, defined as the vector that maximizes the thrust magnitude $T$~\cite{Brandt:1964sa, Farhi:1977sg}:
\begin{equation}
    T = \frac{\sum_i|{{\bf p}_i} \cdot \hat{{\bf n}}_{\rm T}|}{\sum |{\bf p}_i|},
\end{equation}
where ${\bf p}_i$ is the momentum of the $i$-th particle in the CMS. 
To enhance the selection of $(\tau^{+}\to\pi^{+}\pi^{-}\pi^{+}\bar{\nu}_{\tau})(\tau^{-}\to e^{-} + anything)$, 
three-prong candidates are reconstructed in combinations of three pions on the same side of a plane perpendicular to the thrust axis, while the one-prong candidate requires one electron on the opposite side.    

\begin{figure}
\includegraphics[width=8.5cm]{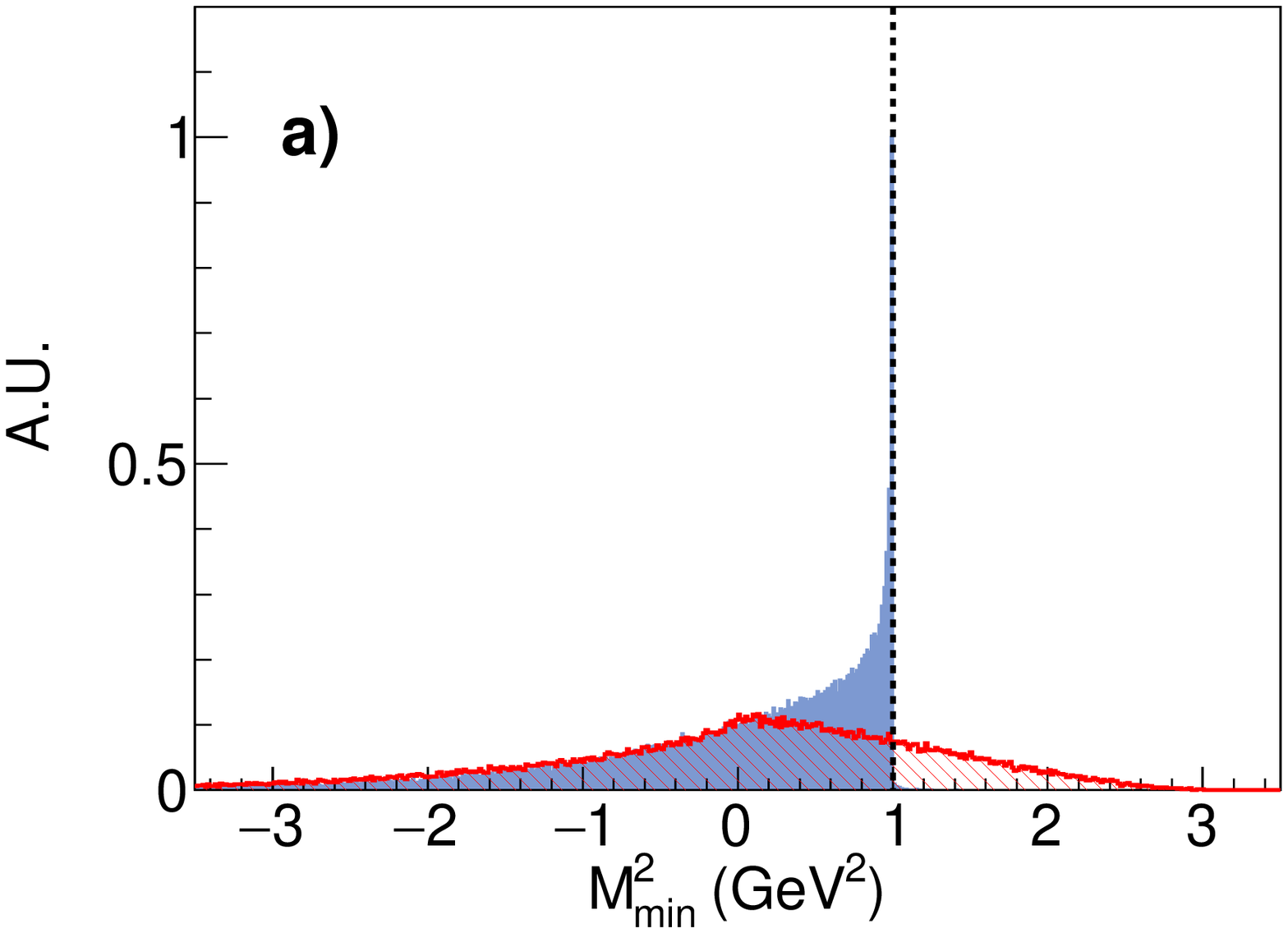}\\
\includegraphics[width=8.5cm]{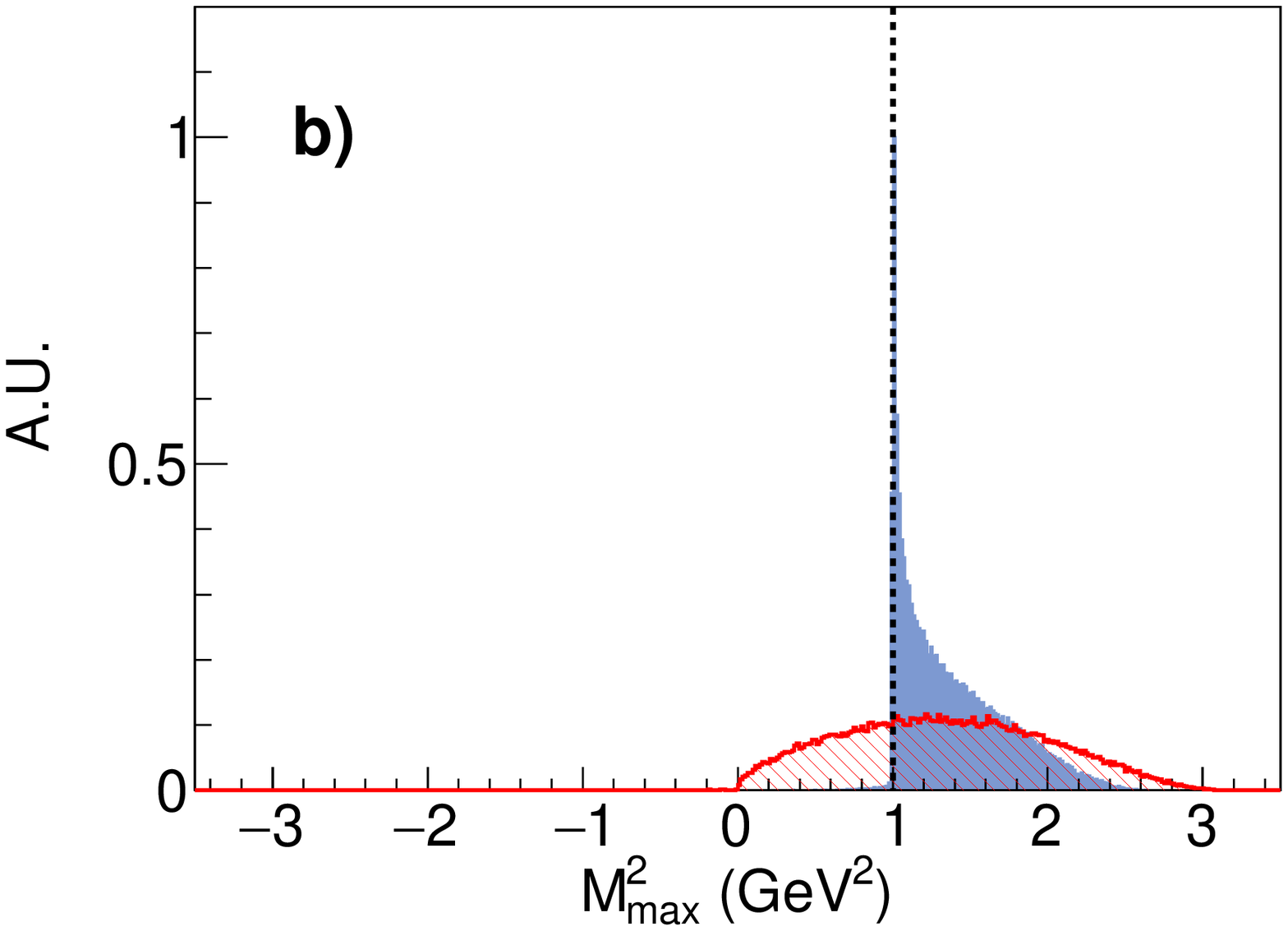}
\caption{ Distribution of (a) $M^2_{min}$  and (b) $M^2_{max}$ in simulated data.
The solid region represents the signal process $(\tau^{+}\to \pi^{+}\pi^{-}\pi^{+} \bar{\nu}_{\tau})(\tau^{-}\to e^{-}  \alpha)$, for $m_{\alpha}=1.0$ GeV, indicated by the black dashed line. The red dashed region shows backgrounds from $\tau^{+}\tau^{-}$ into $(\tau^{+}\to \pi^{+}\pi^{-}\pi^{+}  \bar{\nu}_{\tau})(\tau^{-}\to e^{-} + anything)$ and $q\bar{q}$ pairs similarly reconstructed. For illustration, the signal production is set to an equal number of background events. \label{Mass3x1Alpha2}}
\end{figure}

\begin{figure}
\includegraphics[width=8.5cm]{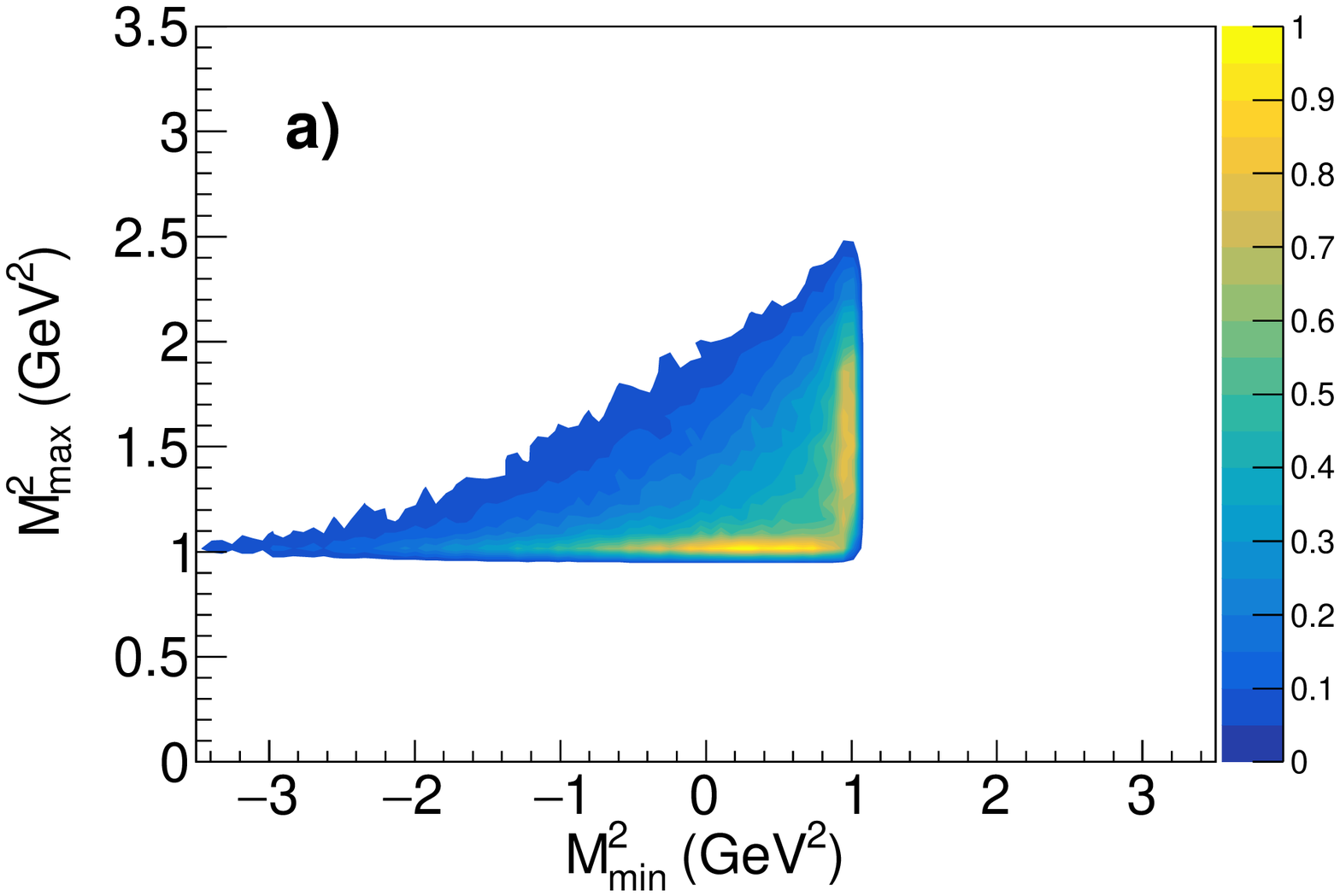}\\
\includegraphics[width=8.5cm]{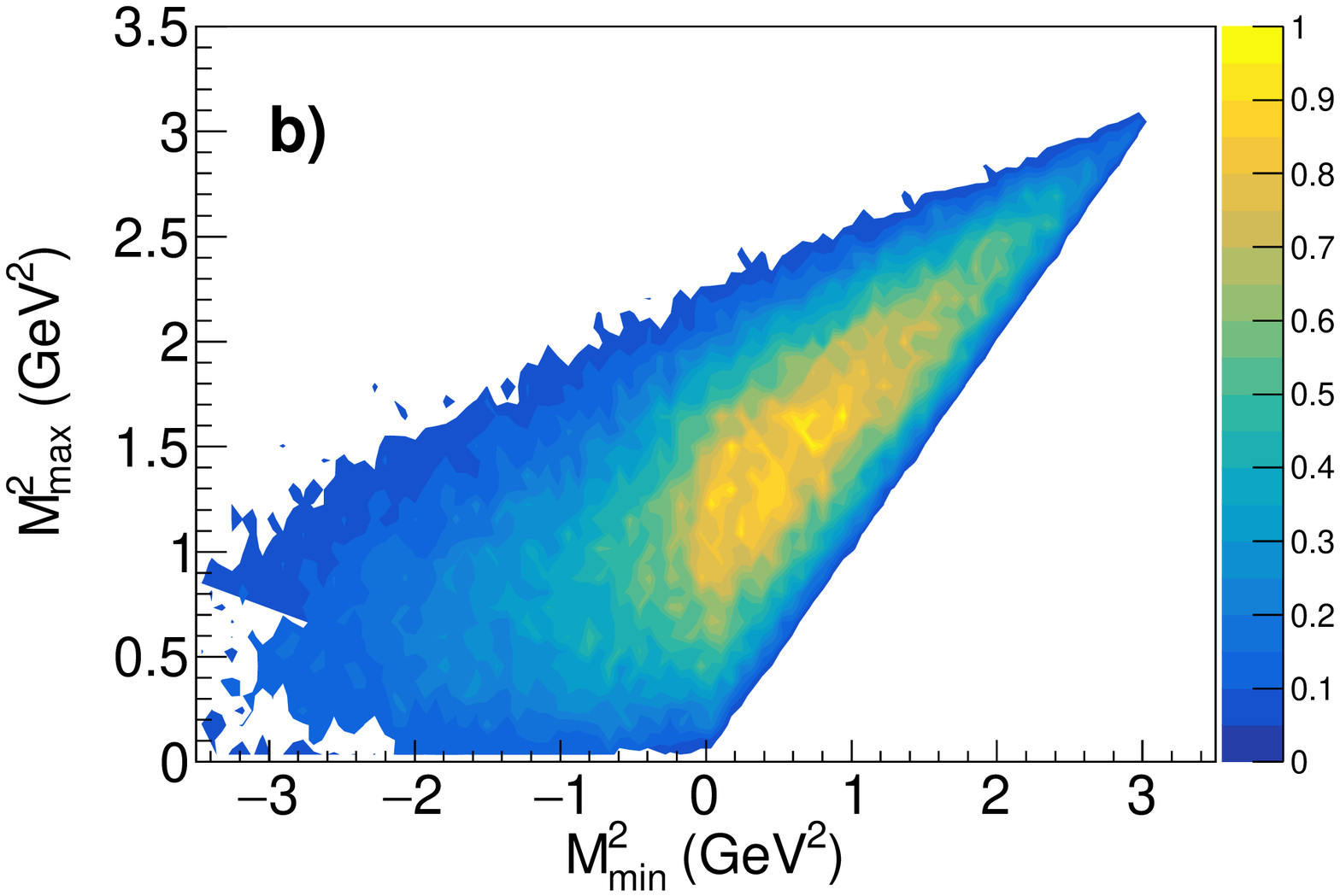}
\caption{ Normalized distribution of $M^2_{min}$  vs $M^2_{max}$ in simulated data for a) signal process $(\tau^{+}\to \pi^{+}\pi^{-}\pi^{+} \bar{\nu}_{\tau})(\tau^{-}\to e^{-}  \alpha)$ with $m_{\alpha}=1.0$ GeV, and b) SM background processes $(\tau^{+}\to \pi^{+}\pi^{-}\pi^{+}  \bar{\nu}_{\tau})(\tau^{-}\to e^{-} + anything)$ and $q\bar{q}$ pairs. For illustration, the signal production is set to an equal number of background events. \label{2DSignalandBkg}}
\end{figure}

Figure \ref{Mass3x1Alpha2} and Fig. \ref{2DSignalandBkg} show the distributions for $M_{min}^2$ and $M_{max}^2$ in the simulated data before momentum smearing. The number of $\tau\to e\alpha$ decays for $m_{\alpha}=1$ GeV has been set equal to the number of SM background events as an example. In both variables, the signal distribution shows clear endpoints, and a peaking structure at $m^{2}_{\alpha}$, and the background extends all over the kinematic allowed region without significantly peaking at any point. These striking differences between signal and background data distributions will allow us to disentangle one from each other.
In the two-dimensional distribution of $(M_{min}^{2}, M_{max}^{2})$, we do not observe a direct correlation for the signal events. However, background events appear to be correlated.

\subsection{Production measurement}

The production of a BSM decay is usually measured relative to a similar SM process; this to cancel out systematic effects related to luminosity, cross-section, and branching ratios.  The $\tau\to e + anything$ data is dominated by $\tau\to e  \bar{\nu}_{e}  \nu_{\tau}$ decays, and this is used as the normalization process in the $\tau\to \ell  \alpha$ production measurement. Then, to estimate this relative production we need to identify three components in the $\tau\to e + anything$ data: the $\tau\to e  \alpha$ decays; the SM process $\tau\to e  \bar{\nu}_{e}  \nu_{\tau}$; and anything else is considered as background. For this, the data should follow a probability distribution given by
\begin{eqnarray}
    f(x) &=& \frac{N_{\alpha} S_{\alpha}({\bf x }) + N_{\nu} S_{\nu}({\bf x}) + N_b B({\bf x })}{N_{\alpha} + N_{\nu} + N_{b}}, \nonumber\\
    &=&  
    \frac{N_{\nu} \mu \frac{\epsilon_{\alpha}}{\epsilon_{\nu}} S_{\alpha}({\bf x}) + 
    N_{\nu} S_{\nu}({\bf x}) +
    N_{b} B({\bf x})}{N_{\nu} \mu \frac{\epsilon_{\alpha}}{\epsilon_{\nu}} + N_{\nu} + N_{b}}, \label{Pdf}
\end{eqnarray}
where $N_{\alpha}$, $N_{\nu}$, and $N_b$ are number of $\tau\to e \alpha$ decays, the number of $\tau\to e  \bar{\nu}_{e} \nu_{\tau}$ decays, and the number of background events, respectively. These components are described by the probability density functions $S_{\alpha}({\bf x})$, $S_{\nu}({\bf x})$ and $B({\bf x})$. Here $\epsilon_{\alpha} /\epsilon_{\nu}$ is the relative observation efficiency of the first two components, and $\mu$ is the relative branching ratio 
\begin{eqnarray}
\mu&=& \frac{Br(\tau\to e  \alpha)}{Br(\tau\to e  \bar{\nu}_{e}  \nu_{\tau})}.    
\end{eqnarray}
This is the parameter of interest in which a non-zero value indicates the presence of a signal in data. Then the measurement of the $\tau\to e \alpha $ production reduces to estimate the value of $\mu$.
For the search of tiny signals, it is better to formulate the $ \mu $ determination in terms of a hypothesis test to exclude a possible signal at the desired confidence level (CL).

To test the performance of the $M_{min}$, $M_{max}$, and 2D methods in the determination of $\mu$, we use simulated data samples of $\tau\to e + anything$ composed of SM-only processes that follow the selection criteria described in Section \ref{Selection}. Then by using the model in Eq. \ref{Pdf}, 95\% C.L. upper limits on $\mu$ are estimated with an asymptotic CLs technique \cite{Cowan:2010js} implemented in RooStats \cite{Moneta:2010pm}. For the data modeling, the probability density distributions $S_{\alpha}({\bf x})$, $S_{\nu}({\bf x})$ and $B({\bf x})$, are extracted as templates from independently simulated data samples, where the relative efficiency is found to be $\epsilon_{\alpha} /\epsilon_{\nu}=1.17$.
Four methods are studied for the upper limit estimate,
\begin{enumerate}
    \item The ARGUS method, using the normalized electron energy in the pseudo-rest-frame, $x=2E_{e}/m_{\tau}$, as the discriminating variable.
    \item The $M_{min}$ method, using $M_{min}^{2}$ as the discriminating variable.
    \item The $M_{max}$ method, using $M_{max}^{2}$ as the discriminating variable.
    \item The 2D method. A combination of $M_{min}^{2}$ and $M_{max}^{2}$ in a two-dimensional density distribution.
\end{enumerate}

\begin{figure}
\includegraphics[width=8.5cm]{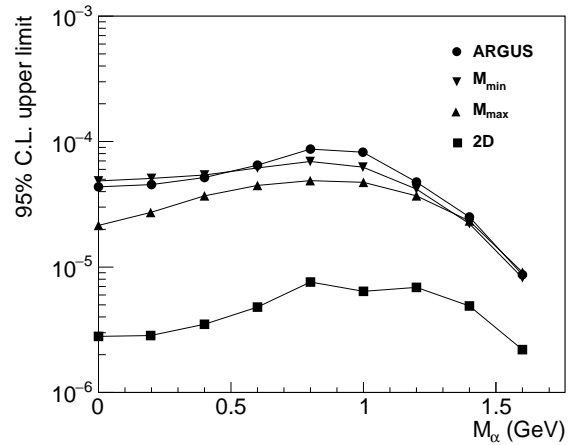}
\caption{ 95\% C.L. upper limits on the relative branching ratio $Br(\tau\to e \alpha)/Br(\tau\to e  \bar{\nu}_{e} \nu_{\tau})$ for an integrated luminosity of 50 ab$^{-1}$ for tau pairs in 3x1 prong decays. \label{LimitsAlpha}}
\end{figure}

Figure \ref{LimitsAlpha} summarizes the results on the upper limit estimate for masses of the $\alpha$ particle between $0$ and $1.6$ GeV for an integrated luminosity of 50 ab$^{-1}$; the data Belle II expects to collect during the next decade. The ARGUS and the $M_{min}$ methods present similar performance for the upper limit estimate. However, for lower $m_{\alpha}$ values, we obtain better upper limits when using the $M_{max}$ variable than with these two methods. This improvement is not negligible at all; for $m_{\alpha}=0$, the upper limit in the $M_{max}$ method is half the one achieved with the ARGUS technique. If we use a simple scaling of $1/\sqrt{N}$ for the limit estimate as data increase, this translates to four times more data in the ARGUS or the $M_{min}$ method to perform as good as the $M_{max}$ variable. However, the 2D method produces a better upper limit than the other three methods alone, improving the expected upper limit by one order of magnitude compared to the ARGUS technique. For the $m_{\alpha}=0$, the ARGUS method upper limit is 15 times larger than the upper limit in the 2D method. Using simple data scaling means the ARGUS technique requires 225 times more data to perform as the 2D method.

\begin{figure}
\includegraphics[width=8.5cm]{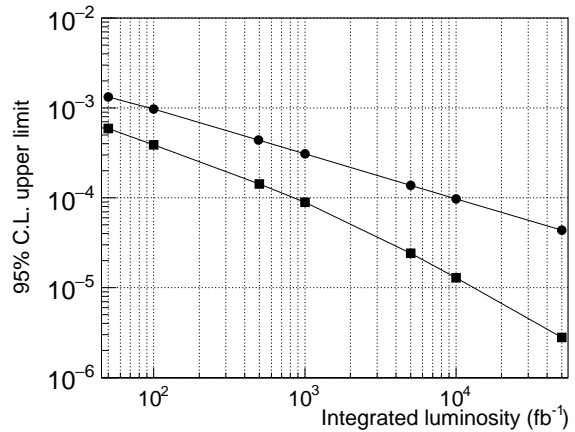}
\caption{ 95\% C.L. upper limits on the relative branching fraction $Br(\tau\to e \alpha)/Br(\tau\to e \bar{\nu}_{e} \nu_{\tau})$ as a function of the integrated luminosity for tau pairs in 3x1 prong decays. Black circle (squared) points are for ARGUS (2D) method. The upper limit are for $m_{\alpha}=0$. \label{LimitsAlphaLuminosity}}
\end{figure}

For $m_{\alpha}=0$, Fig.  \ref{LimitsAlphaLuminosity} shows for the ARGUS and the 2D method the upper limit on the relative branching ratio as a function of the integrated luminosity. We note that to reach an upper limit below $10^{-4}$, 1 ab $ ^{-1}$ of data is necessary for the 2D method. However, we would require an order of magnitude more statistics for the ARGUS technique to reach this precision.
With the proposed 2D method, the Belle II experiment could reach the level of the upper limit in the ARGUS technique for the full data sample, but with only a fraction of the data, which could be collected during the first years of operation.

\section{Conclusions}

We have studied the kinematics of the decay of a particle-antiparticle pair for known center-of-mass energy when in each decay, one of the produced particles escapes detection. This study led us to determine kinematic bounds on the mass of the new $\alpha$ particle in the search for LFV $\tau\to \ell \alpha$ decays. We propose using these bounds in a two-dimensional method for the production measurement of this BSM process in tau pair decays at electron-positron colliders.  

For the upper limit estimate on the relative production of the $ \alpha $ particle, we apply the method to simulated data that emulates some of the Belle II experiment conditions. The proposed variables, $M_{min}$ and $M_{max}$, show similar or better performance than the commonly used ARGUS technique.  When we combine both variables in the 2D method, the upper limit estimate is one order of magnitude lower than the one obtained by the ARGUS, $M_{min}$ or $M_{max}$ methods alone. 
With the 2D method, Belle II could attain the expected upper limit in the ARGUS technique with only a fraction of the full data sample to be collected during its operation.
 
For the performance comparison of the methods presented in this work, the upper limit estimate lacks several experimental effects, such as trigger efficiencies, beam backgrounds, or particle identification efficiency. However, in most cases, these effects will cancel out in the ratio of signal to background processes or be a global scale factor in the data distribution model, and they will similarly affect any of the methods considered.  More importantly, they should not change the relative performance of the discriminant variables, in which the 2D approach achieves the best upper limit estimate. 

One important difference between the proposed approach and the ARGUS method is that we do not need a three-prong tau decay as required in the pseudo-rest-frame technique, then our methods can be implemented in 1x1 prong decays such as $(\tau^{+}\to\pi^{+}\bar{\nu}_{\tau})(\tau^{-}\to e^{-}\alpha)$, or any $n$x1 prong tau decay. Since $Br(\tau^{-}\to\pi^{-}\nu_{\tau})= 1.16 \times Br(\tau^{-}\to\pi^{-}\pi^{+}\pi^{-}\nu_{\tau})$ \cite{PDG2018}, we expect an upper limit of the same order of magnitude for 1x1 prong decays as the obtained from 3x1 prong decays, and combined will increase even more the reach of Belle II on the production search for LFV $\tau\to e \alpha$ decays. 

Although it is beyond the scope of the present work, we should mention that the proposed methods could be applied to studies on the tau neutrino mass upper limit from colliders \cite{Barate:1997zg}, or for heavy neutrinos searches in the tau lepton sector \cite{Kobach:2014hea}. Also, when in Eq. \ref{EqMaster} we take $m_{X}=m_{\tau}$ as the parameter of interest, and SM processes are required in the two tau decays, endpoints can be found for the mass measurement of the tau lepton.

\section*{Acknowledgements} \label{sec:acknowledgements}

We wish to thanks P. Roig, I. Heredia de la Cruz, and H. Castilla Valdez for helpful discussions. This work was supported by the SEP-CINVESTAV research grant 237.

\bibliography{references.bib} 

\end{document}